\documentclass[aps,prd]{revtex4}
\usepackage{epsf}
\usepackage{graphicx}
\usepackage{subfigure}
\usepackage{amssymb}
\usepackage{color}
\newcommand{\be}{\begin{equation}}
\newcommand{\ee}{\end{equation}}
\newcommand{\ba}{\begin{eqnarray}}
\newcommand{\ea}{\end{eqnarray}}
\newcommand{\no}{\nonumber\\}
\newcommand{\ra}{\rightarrow}
\newcommand{\ie}{{\it i.e.}\ }

\newcommand{\grts}{\raise.3ex\hbox{$>$\kern-.75em\lower1ex\hbox{$\sim$}}}
\newcommand{\lets}{\raise.3ex\hbox{$<$\kern-.75em\lower1ex\hbox{$\sim$}}}

\begin{document}
\title{Heavy Higgs Searches and  Constraints on Two Higgs Doublet Models}
\author{Chien-Yi Chen and Sally Dawson}

\affiliation{
Department of Physics, Brookhaven National Laboratory, Upton, New York, 11973}

\author{Marc Sher}
\affiliation{High Energy Theory Group, College of William and Mary, Williamsburg, Virginia 23187, U.S.A.}

\date{\today}

\begin{abstract}
Since the discovery of a Higgs boson at the LHC and the measurement of many of its branching ratios, there have been numerous studies exploring the restrictions these results place on the parameter space of two Higgs doublet models.  
 We extend these results  to include the full data set and study the expected sensitivity that can be obtained with 
  $300~$fb$^{-1}$ and $3000$~fb$^{-1}$ integrated luminosity.
 We  consider searches for a heavy Standard Model Higgs boson, with a mass ranging from $200$ to $400$ GeV, 
 and show that the  non-observation of such a Higgs boson can substantially narrow the allowed regions of parameter space
 in two Higgs doublet models.
\end{abstract}

\maketitle

\section{Introduction}

Following the discovery of a Higgs boson, experiments at the LHC 
can begin to probe the electroweak symmetry breaking sector.     
Their task is  to measure the properties of the Higgs boson as precisely as possible.   
Any deviation from the Standard Model predictions would be evidence of physics beyond the Standard Model.        
Many extensions of the Standard Model have been proposed over the past few decades, and many contain an electroweak symmetry breaking sector with more than one Higgs doublet.    These extensions can easily accommodate a $125$ GeV scalar, but also typically 
predict deviations in its couplings.     
Thus, it is crucial to examine extensions of the Standard Model and determine the expectations for the couplings of the $M_{h^0}=125$ GeV scalar.       Some
 of the simplest extensions of the scalar sector are the two Higgs doublet models (2HDMs)    
\cite{Branco:2011iw}.     
The 2HDMs contain five physical Higgs scalars:   a charged Higgs $H^\pm$, a pseudoscalar $A$,
 and two neutral scalars, $h$ and $H$.   Although it is possible that the $125$ GeV state is the heavier of the neutral scalars \cite{Ferreira:2012my,Drozd:2012vf,Chang:2012ve}, we  assume here that it is the lighter.

In general, 2HDMs have Higgs mediated tree level flavor changing neutral currents (FCNCs), 
which must be suppressed.   Most 2HDMs eliminate FCNCs by imposing a discrete $Z_2$ symmetry in which the 
fermions of a given charge only couple to one of the Higgs doublets.    The two most familiar versions are the type I model, in which all of the fermions couple to the same Higgs doublet, and the type II model, in which the $Q=2/3$ quarks couple to one doublet and the $Q=-1/3$ quarks and leptons couple to the other.    
Two additional versions interchange the lepton assignments.  In the ``lepton-specific" model, all of the quarks couple to one doublet while the leptons couple to the other, and in the ``flipped" model, the $Q=2/3$ quarks and leptons couple to one doublet and the $Q=-1/3$ quarks couple to the other.    All four of these models have been extensively studied \cite{Branco:2011iw}.   
The couplings of the Higgs bosons to fermions are described by two free parameters.   
The ratio of vacuum expectation values of the two Higgs doublets is  $\tan\beta \equiv \frac{v_2}{v_1}$, 
and the mixing angle which diagonalizes the neutral scalar mass matrix is $\alpha$.   The couplings of the light (heavy) CP even Higgs boson, 
$h^0$ ($H^0$), to fermions and gauge bosons
relative to the Standard Model couplings
 are given for all four 2HDMs considered here in Table \ref{table:coups} (Table \ref{table:coupsH}). 

\begin{table}[t]
\caption{Light Neutral Higgs ($h^0$) Couplings in the 2HDMs}
\centering
\begin{tabular}[t]{|c|c|c|c|c|}
\hline\hline
& I& II& Lepton Specific& Flipped\\
\hline
$g_{hVV}$ & $\sin(\beta-\alpha)$ & $\sin (\beta-\alpha)$ &$\sin (\beta-\alpha)$&$\sin (\beta-\alpha)$\\
$g_{ht\overline{t}}$&${\cos\alpha\over\sin\beta}$&${\cos\alpha\over\sin\beta}$&${\cos\alpha\over\sin\beta}$&${\cos\alpha\over\sin\beta}$\\
$g_{hb{\overline b}}$ &${\cos\alpha\over\sin\beta}$&$-{\sin\alpha\over\cos\beta}$&${\cos\alpha\over\sin\beta}$&$-{\sin\alpha\over \cos\beta}$\\
$g_{h\tau^+\tau^-}$&${\cos\alpha\over \sin\beta}$&$-{\sin\alpha\over \cos\beta}$&$-{\sin\alpha\over \cos\beta}$&${\cos\alpha\over\sin\beta}$\\
\hline
\end{tabular}
\label{table:coups}
\end{table}

\begin{table}[t]
\caption{Heavy Neutral CP Even Higgs ($H^0$) Couplings in the 2HDMs}
\centering
\begin{tabular}[t]{|c|c|c|c|c|}
\hline\hline
& I& II& Lepton Specific& Flipped\\
\hline
$g_{HVV}$ & $\cos(\beta-\alpha)$ & $\cos (\beta-\alpha)$ &$\cos (\beta-\alpha)$&$\cos (\beta-\alpha)$\\
$g_{Ht\overline{t}}$&${\sin\alpha\over\sin\beta}$&${\sin\alpha\over\sin\beta}$&${\sin\alpha\over\sin\beta}$&${\sin\alpha\over\sin\beta}$\\
$g_{Hb{\overline b}}$ &${\sin\alpha\over\sin\beta}$&${\cos\alpha\over\cos\beta}$&${\sin\alpha\over\sin\beta}$&${\cos\alpha\over \cos\beta}$\\
$g_{H\tau^+\tau^-}$&${\sin\alpha\over \sin\beta}$&${\cos\alpha\over \cos\beta}$&${\cos\alpha\over \cos\beta}$&${\sin\alpha\over\sin\beta}$\\
\hline
\end{tabular}
\label{table:coupsH}
\end{table}

Following the initial evidence for a Higgs boson
 at $ M_{h^0}=125$ GeV,
  Ferraira, et al\cite{Ferreira:2011aa}  studied the implications of such a Higgs particle
 for the four versions of the 2HDMs and
 presented
  the expected branching ratios of the $M_{h^0}=125$ GeV state.   
  Subsequently,
   many papers
   \cite{Chen:2013kt,Alves:2012ez,Craig:2012vn,Craig:2012pu,Bai:2012ex,Azatov:2012qz,Dobrescu:2012td,Ferreira:2012nv,Celis:2013rcs,Grinstein:2013npa,Krawczyk:2013gia,Coleppa:2013dya,Altmannshofer:2012ar,Chiang:2013ixa,Basso:2012st,Chang:2012zf,Craig:2013hca} 
  examined various channels in the four 2HDMs in light of the experimental findings at the LHC.   
  More recently,  \cite{Barroso:2013zxa} updated the study of Type I and Type II models,  using the entire 
  LHC dataset.
  In Section II, we update previous studies \cite{Chen:2013kt} for all four 2HDMs to include the full data set and highlight the
  significant effect of the latest CMS result on $h\rightarrow \gamma\gamma$ on the global fit.  We
extend previous results to demonstrate the expected sensitivity with  $300$~fb$^{-1}$ or $3000$~fb$^{-1}$.   
In Section III,  we show that current ATLAS and CMS bounds on a ${\it heavy}$ Higgs boson, with mass
between 200 and 400 GeV, can 
bound regions of parameter space that have not yet been covered by 
the
analysis of the $M_{h^0}=125$ GeV Higgs decays, 
and we extend these limits as well to $300~$fb$^{-1}$  and $3000~$fb$^{-1}$.

\section{LHC Reach from $h^0$ Measurements}

Previous analyses examined individual decays of the $M_{h^0}=125$ GeV 
Higgs, the $h^0$, and looked at the implications for 2HDMs, finding the regions in the $(\alpha,\beta)$ parameter-space allowed by current LHC data.    Ref. \cite{Chen:2013kt}
determined, for each of the four 2HDMs, the allowed regions of parameter-space.
We have updated their results to include the most recent experimental
data and have also studied the bounds that can be obtained at a future LHC with 14 TeV and 
 integrated luminosities of $300~$fb$^{-1}$ and
$ 3000~$fb$^{-1}$.   To estimate these bounds, we look at the current errors, assume that the Standard Model 
prediction is correct, and scale the errors as $1/\sqrt{N}$, where $N$ scales like the integrated luminosity.   
This corresponds to 'scheme 2' of the CMS\cite{nisati} high luminosity projections\footnote{ The assumption that the uncertainty scales as $1/\sqrt{L}$ is true for
the statistical error, and assumes that 
improvements are made in the systematic errors with increasing luminosity.  This assumption is therefore optimistic, but is the one used by CMS for
the European Strategy Report\cite{nisati}.}.

A $\chi^2$ fit to the data shown in Tables \ref{tab:models1} and \ref{tab:models2}  is performed assuming 
$M_{h^0}=125$ GeV. We follow the standard definition of $\chi^2 = \Sigma_i {(R_i^{\rm 2HDM}-R_i^{\rm meas})^2\over (\sigma^{meas}_i)^2}$, where $R^{2HDM}$ 
represents predictions for the signal strength from the 2HDMs and $R^{\rm meas}$ stands for the measured 
signal strength shown in Tables \ref{tab:models1} and \ref{tab:models2}. When the errors are asymmetric, we have averaged them in quadrature,
$\sigma=\sqrt{
{(\sigma_+)^2 +(\sigma_-)^2\over 2}}$.  Although including the asymmetric errors in the analysis would in general provide more accurate information, in this case
the only data with substantial asymmetric errors are the CMS vector boson fusion channel with $h^0\rightarrow ZZ$ and the Tevatron gluon fusion channel
with $h^0\rightarrow \tau^+\tau^-$, both of which have relatively little pull on the overall $\chi^2$.  Therefore this
assumption will have only a very minor effect on our results.

\begin{figure}[tb]
\subfigure[]{
      \includegraphics[width=0.36\textwidth,angle=0,clip]{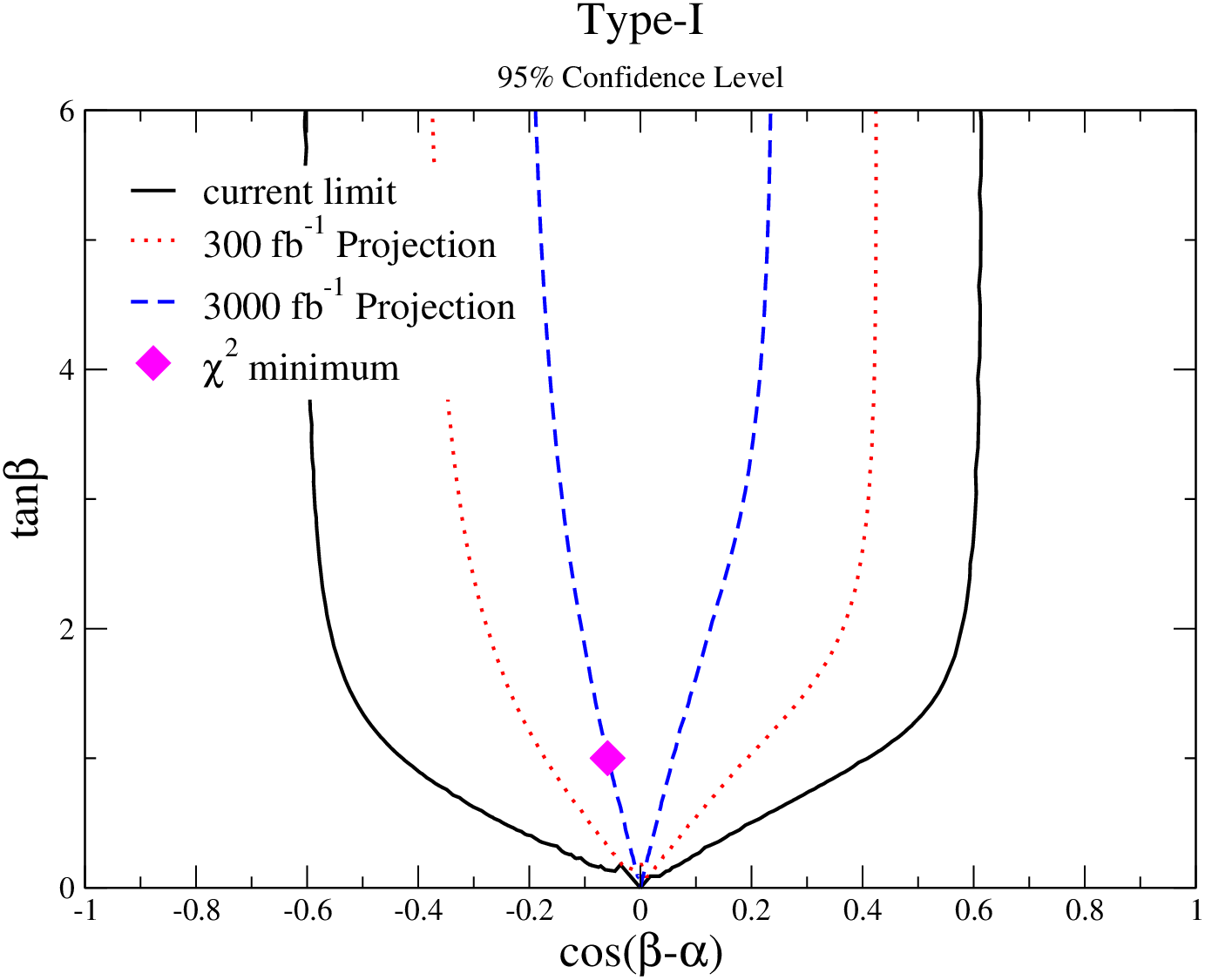}
}
\subfigure[]{
      \includegraphics[width=0.36\textwidth,angle=0,clip]{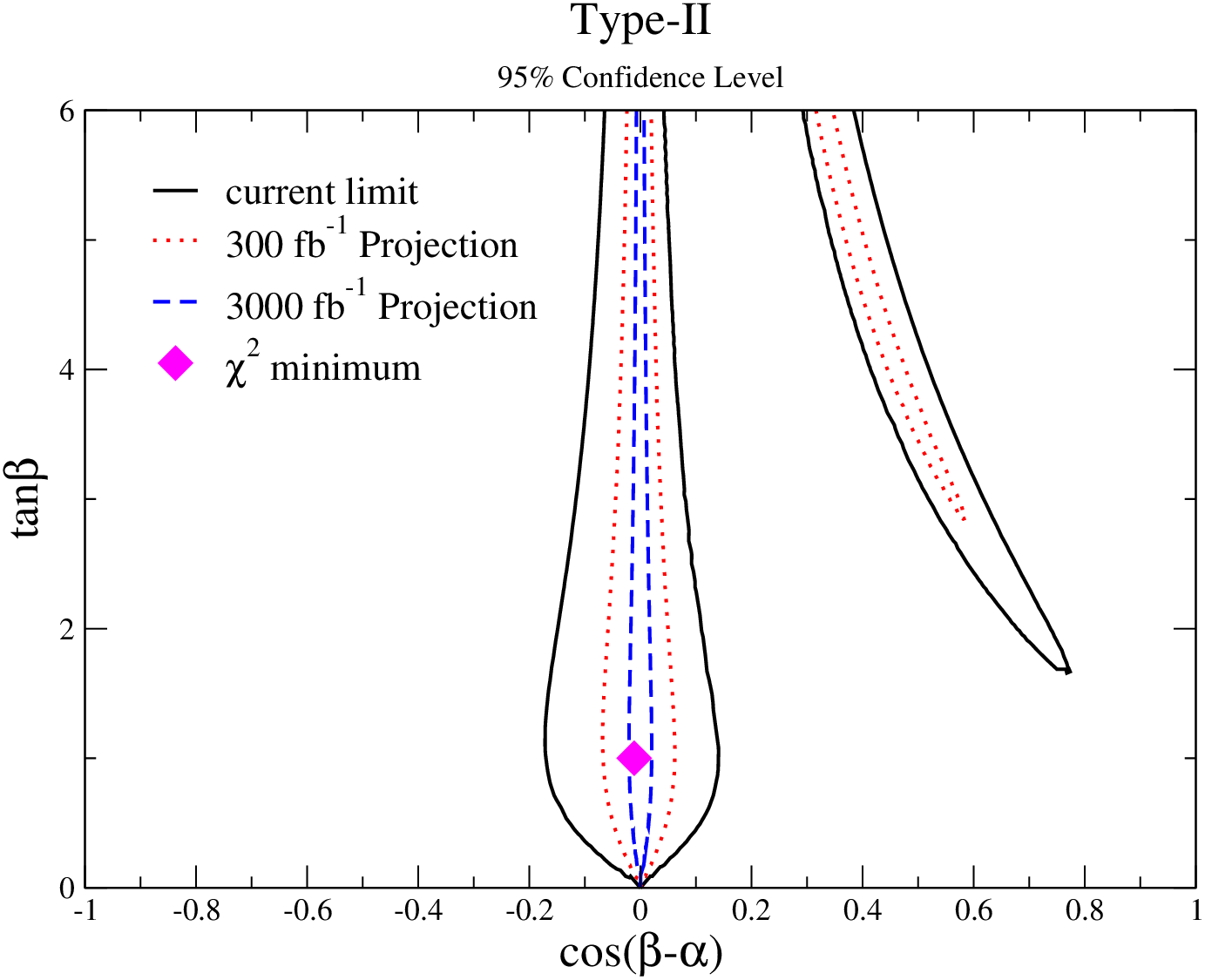}
}
\subfigure[]{
      \includegraphics[width=0.36\textwidth,angle=0,clip]{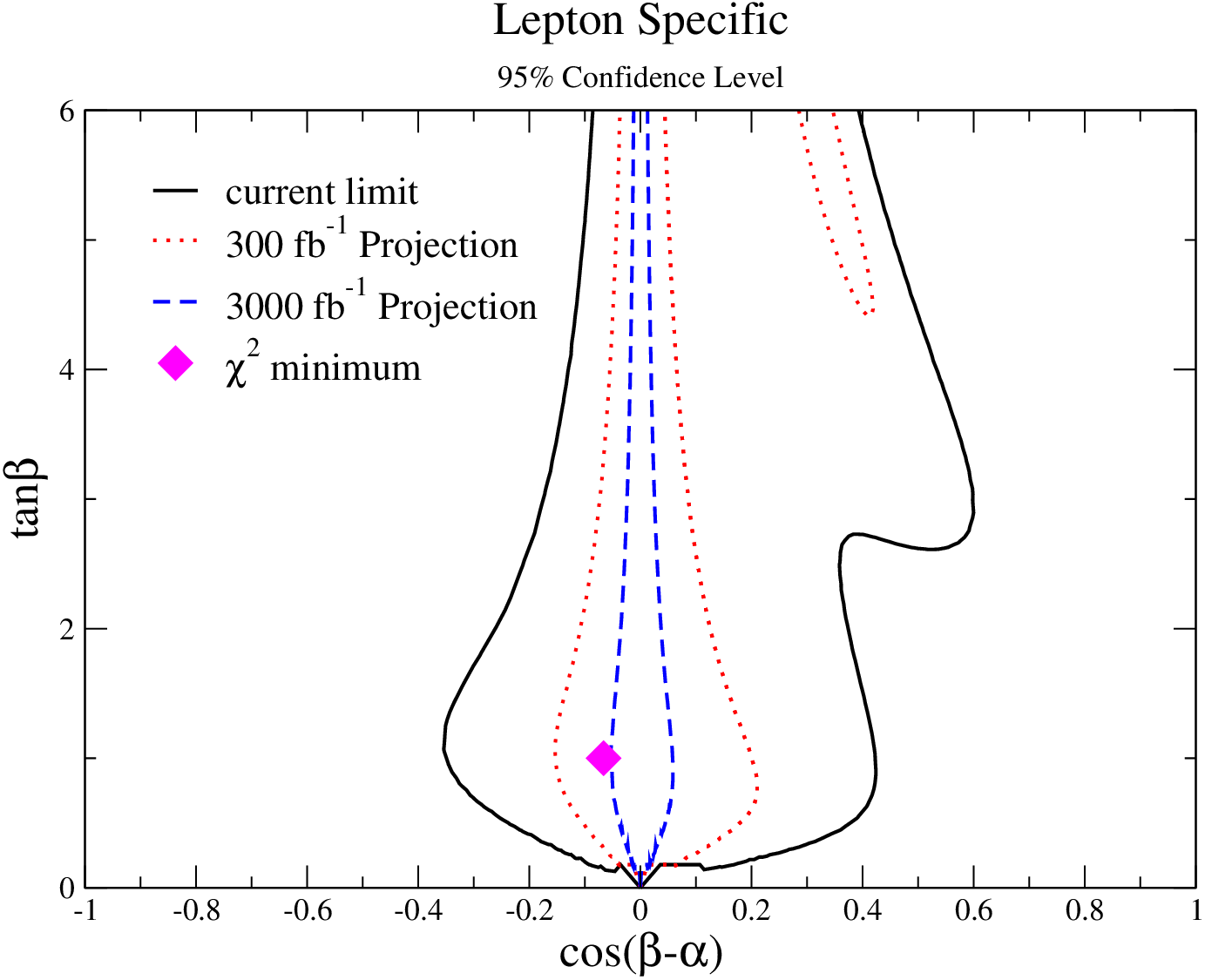}
}
\subfigure[]{
      \includegraphics[width=0.36\textwidth,angle=0,clip]{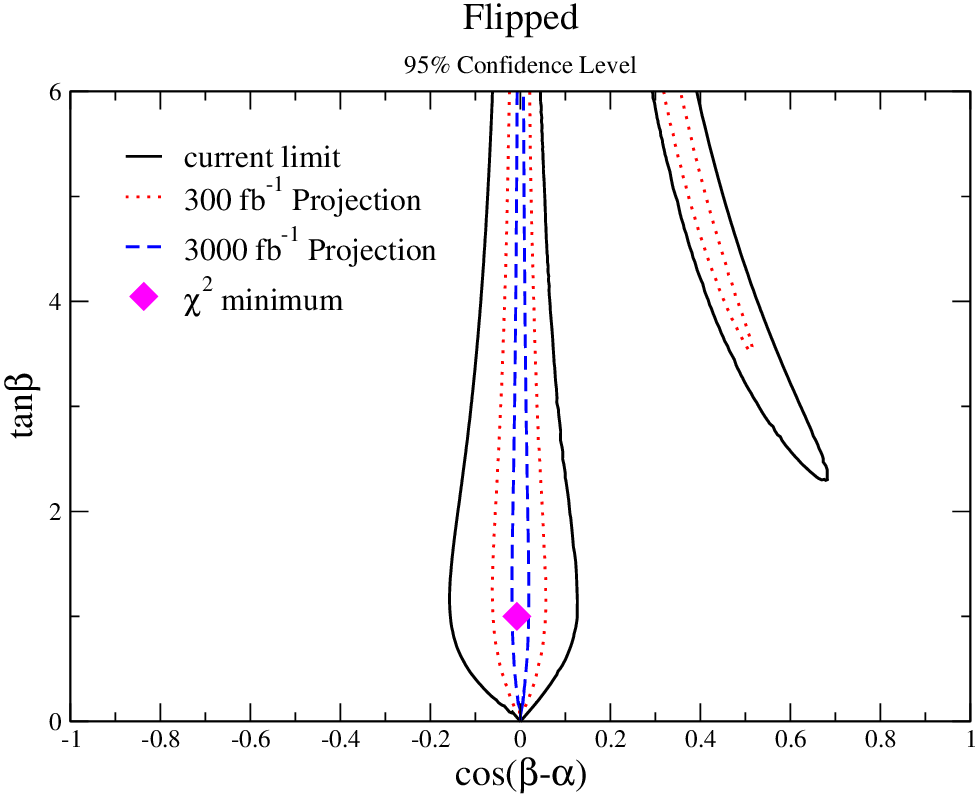}
}
\caption{Allowed regions  in the $(\cos(\beta-\alpha),\tan\beta)$ plane  in Type I (a), Type II (b), Lepton Specific (c),
and Flipped (d) 2HDMs obtained by performing  a $\chi^2$ analysis.  
The region between the black (solid), red (dotted), and blue (dashed)
 lines is allowed at $95\%$ confidence level corresponding to the current limits and the projected limits for integrated luminosities of 
 $300~$fb$^{-1}$ and $3000~ $fb$^{-1}$, respectively.
}
\label{chisq_fig}
\end{figure}

Our results are given in Fig. \ref{chisq_fig}. For each of the four models, we plot the current limits on the parameter-space, and the
 projected limits for integrated luminosities of $300~$fb$^{-1}$ and $3000~$fb$^{-1}$\footnote{We use the same sign convention as in 
 Ref \cite{Gunion:1989we},  \ie sin$(\beta-\alpha) > 0$ and $0 \leq \beta \leq \frac{\pi}{2}$.}. 
 Bounds from flavor physics  constrain $\tan\beta \ge1$ \cite{Chen:2013kt,Mahmoudi:2009zx}
 and we take this as a prior when we determine the chi-squared minima.
In all of the models the minimum of the $\chi^2$ occurs for $\tan\beta\sim 1$ and $\cos(\beta-\alpha)\sim 0$, 
demonstrating that the couplings of a 2HDM are already constrained
to be close to the Standard Model values.   Similar bounds for the Type-I and Type -II models have been obtained in Ref. \cite{cgt,Craig:2013hca}.
 The parameter-space for the Type-I model is not very constrained at present.  This is because, in the large $\tan\beta$ 
 limit, the Higgs is fermiophobic and production through gluon fusion is suppressed.    Increasing the integrated luminosity will gradually narrowed the allowed parameter-space.   The lepton-specific model is also not severely constrained, because of the enhanced decay to $\tau$ leptons, which is poorly
 measured at present.    For large $\tan\beta$, the bottom-quark Yukawa coupling becomes substantial in the Type-II and flipped models, and thus the 
 currently allowed parameter-space is much more restricted.   
 We do not show  a very small allowed (by LHC data) region in the lower right for $\tan\beta\sim 0-.5$ because that
 region  is excluded  by B physics constraints.  For each of the models considered here, the measured value of $\Delta M_{B_d}$ excludes
 such  small values of $\tan\beta$\cite{Chen:2013kt}.

The most general potential with $2$ Higgs doublets, $\Phi_1$ and $\Phi_2$, and a softly-broken $Z_2$ symmetry is
\begin{eqnarray}
V &=&
m^2_{11}\, \Phi_1^\dagger \Phi_1
+ m^2_{22}\, \Phi_2^\dagger \Phi_2 -
 \mu^2\, \left(\Phi_1^\dagger \Phi_2 + \Phi_2^\dagger \Phi_1\right)
+ \frac{\lambda_1}{2} \left( \Phi_1^\dagger \Phi_1 \right)^2
+ \frac{\lambda_2}{2} \left( \Phi_2^\dagger \Phi_2 \right)^2
\no & &
+ \lambda_3\, \Phi_1^\dagger \Phi_1\, \Phi_2^\dagger \Phi_2
+ \lambda_4\, \Phi_1^\dagger \Phi_2\, \Phi_2^\dagger \Phi_1
+ \frac{\lambda_5}{2} \left[
\left( \Phi_1^\dagger\Phi_2 \right)^2
+ \left( \Phi_2^\dagger\Phi_1 \right)^2 \right].
\end{eqnarray}
As free parameters, one can use the four scalar masses, along with $\alpha$, $\beta$, and $\mu^2$.   In terms of these parameters, 
one finds\cite{Barroso:2013awa},
\begin{equation}
\lambda_1 = \frac{1}{v^2\cos^2\beta}\left( \cos^2\alpha M^2_{H^0} + \sin^2\alpha M^2_{h^0} - \mu^2\tan\beta \right)
\end{equation}
where $v = 246$ GeV.
If one considers the $Z_2$ symmetric case,  
then $\mu^2=0$, and this leads, since $M^2_{H^0} > M^2_{h^0}$, to a lower bound on 
\begin{equation}
\lambda_1 >0.25 (1+\tan^2\beta)\,. 
\end{equation}
 Clearly, for large $\tan\beta$, $\lambda_1$
 becomes non-perturbative.   Requiring  ${\lambda_1\over 4\pi} < 1$
implies $\tan\beta < 7$.  We therefore concentrate on this region  of relatively small $\tan\beta$.     However, if $\mu^2\ne 0$, then 
parameters  can be chosen to avoid this constraint, although some fine-tuning is then required.

\section{Constraints from  Heavy Higgs Searches}

ATLAS and CMS have obtained  upper bounds on a Standard Model Higgs boson
 with a  mass between 150 and 600 GeV and assuming a Standard Model width.   
 We use the $95\%$ confidence level band from recent CMS bounds (from Figure 11 in Ref. \cite{Chatrchyan:2013yoa})
 and  scale predictions as the inverse square root of the integrated luminosity.

For example, suppose $M_{H^0}$ is $200$~GeV.   A Standard Model Higgs boson
 of $200$ GeV will decay almost $100\%$ of the time into vector bosons.  
This is also true (except for extreme values of the parameters) in a 2HDM.    
 The production rate through gluon fusion in the 2HDM will be different than the Standard Model rate
  because of the different $t$ and $b$ couplings.
Thus, the upper bound from ATLAS and CMS on the cross section relative to the Standard Model rate will  
   place a constraint on $\alpha$ and $\beta$. 

\begin{figure}[tb]
\subfigure[]{
      \includegraphics[width=0.36\textwidth,angle=0,clip]{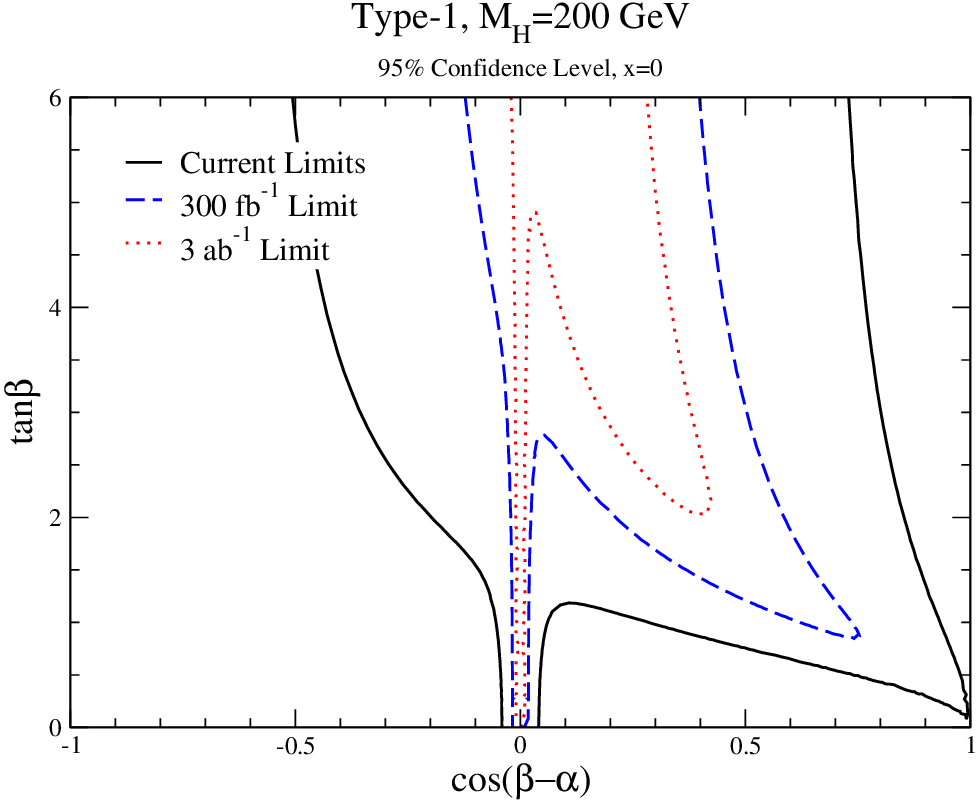}
}
\subfigure[]{
      \includegraphics[width=0.36\textwidth,angle=0,clip]{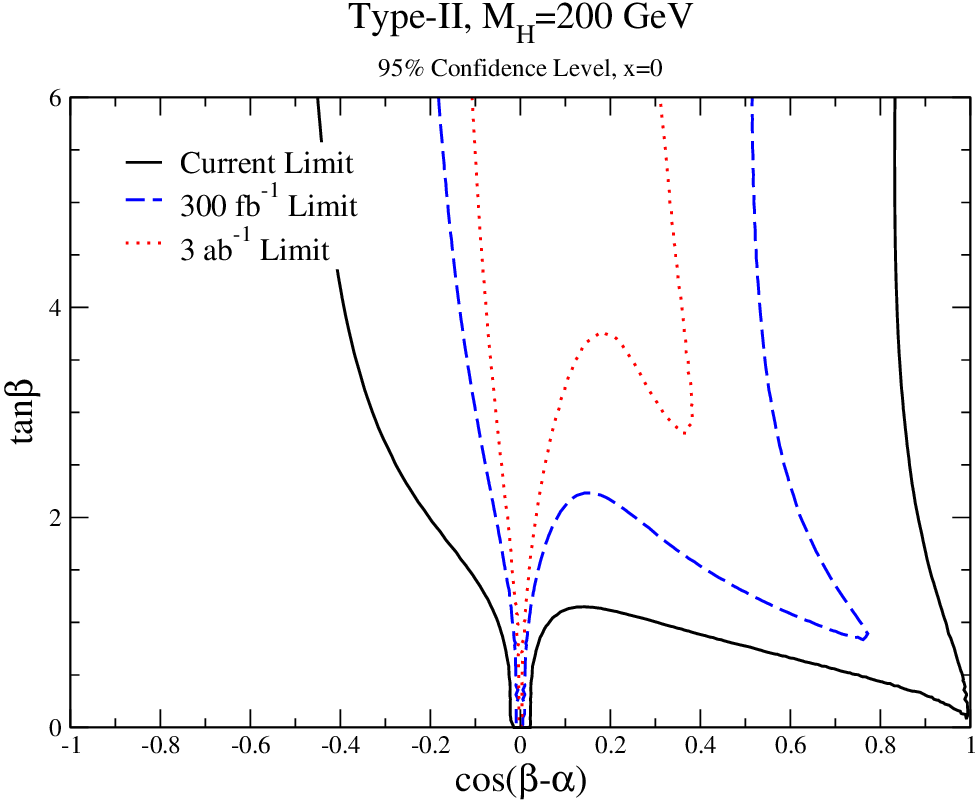}
}
\caption{ Allowed regions in Type-I (a) and Type-II (b) 2HDMs from the LHC limit on a $200~GeV$ heavy Higgs boson.
The region between the black (solid), blue (dashed) and red (dotted) curves is allowed at $95\%$ confidence level corresponding to
the current limits and the projected limits for integrated luminosities of $300~$fb$^{-1}$ and $3000~$fb$^{-1}$, respectively.
}
\label{fg:mh2}
\end{figure}
\begin{figure}[tb]
\subfigure[]{
      \includegraphics[width=0.36\textwidth,angle=0,clip]{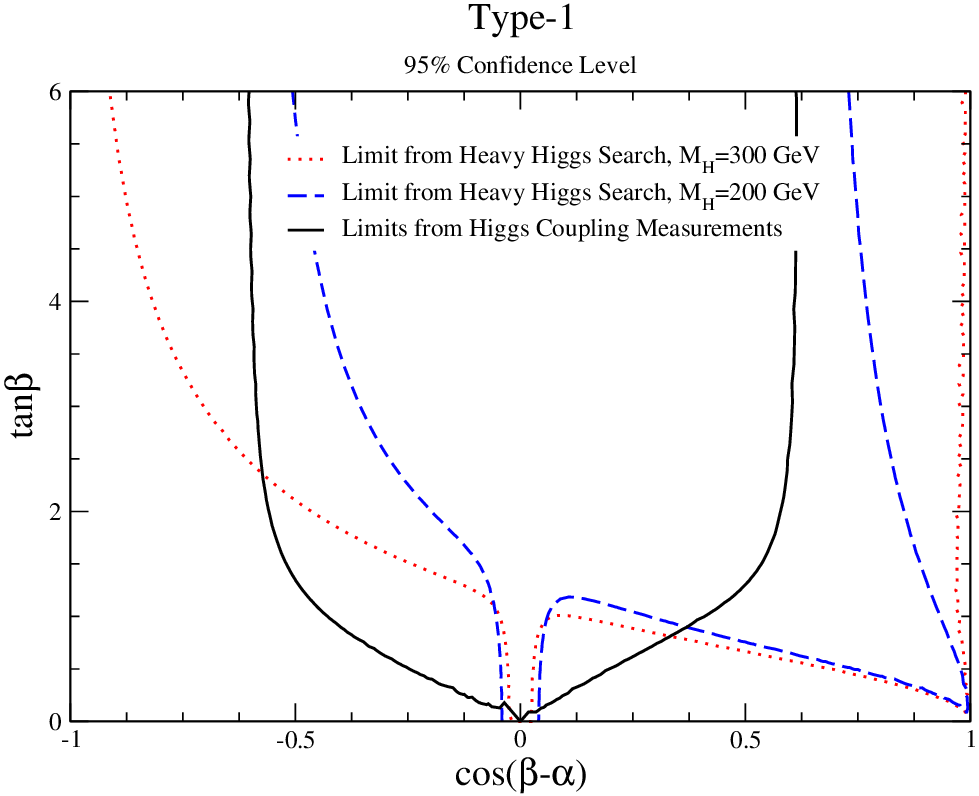}
}
\subfigure[]{
      \includegraphics[width=0.36\textwidth,angle=0,clip]{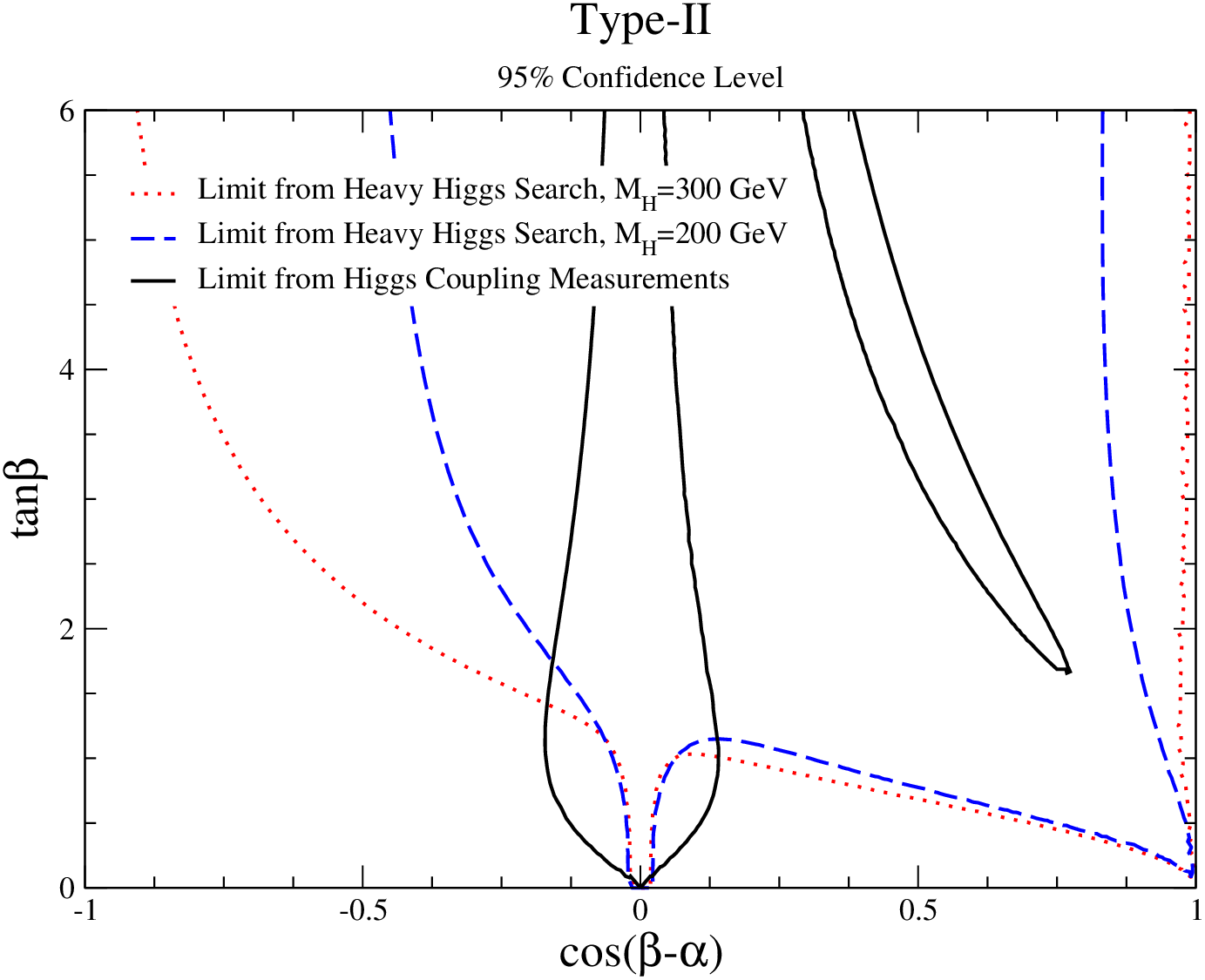}
}
\subfigure[]{
      \includegraphics[width=0.36\textwidth,angle=0,clip]{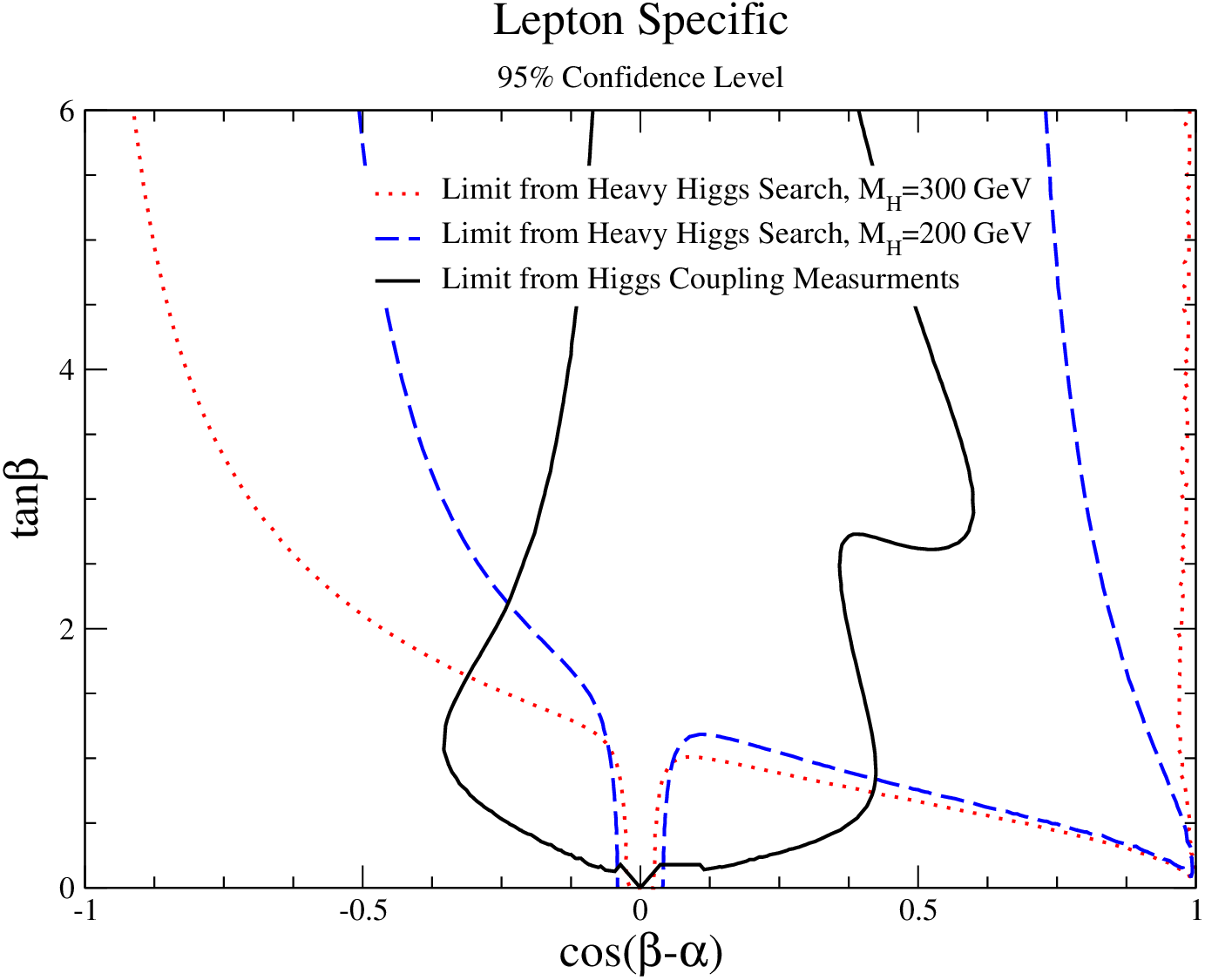}
}
\subfigure[]{
      \includegraphics[width=0.36\textwidth,angle=0,clip]{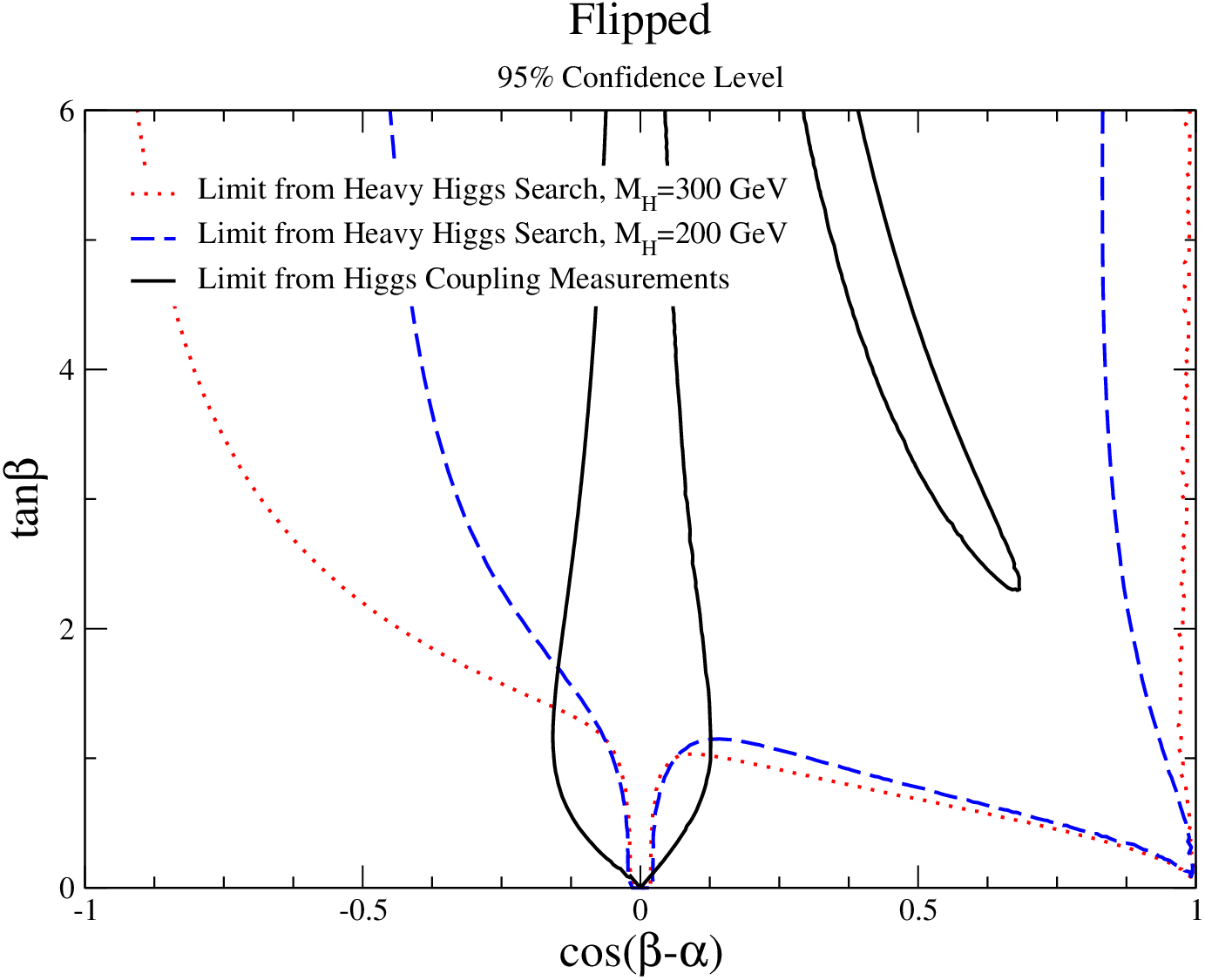}
}
\caption{Allowed regions in Type-I (a), Type-II (b), Lepton Specific (c), and Flipped (d)  2HDMs from the LHC limit on a $200$~GeV 
and a $300$ GeV heavy Higgs boson (blue, red) and the current limits from light Higgs decays (black).
}
\label{fg:MH23}
\end{figure}

For $M_{H^0} = 200$ GeV, we find the results in Fig. \ref{fg:mh2}.
    We show results for the type-I and type-II models, with the current limits and projections for
    $300~$fb$^{-1}$ and
     $3000~$fb$^{-1}$.   
    The lepton-specific and flipped models give very similar results to the type-I and type-II models, respectively. 
    An increase in luminosity will tightly constrain $\cos(\beta-\alpha)$ for $\tan\beta < 4$ in the type-I model 
    and will give a significant constraint for $\tan\beta < 4$ in the type-II model. 
      In Fig. \ref{fg:MH23}, we compare 
current limits from measurements of light Higgs decays with the limits obtained from the heavy Higgs search.  
We see that even with current bounds,
 a significant fraction of the previously allowed parameter-space in the type-I model is excluded by the heavy Higgs search results, and this
  fraction  grows with increasing integrated luminosity (unless, of course, the heavy Higgs is discovered).     For the type-II model, some of the remaining parameter-space is excluded, especially for small $\tan\beta$.   This is a significant result, and shows that the allowed parameter-space of a 2HDM can be substantially narrowed by considering bounds from heavy Higgs searches. 

Once the mass of the  heavy Higgs, $H^0$,
 exceeds $250$ GeV, then  the decay $H^0\rightarrow h^0h^0$  is allowed, which will suppress the branching ratio of the $H$ into vector bosons.  
 The decay width for $H^0\ra h^0h^0$ depends  on $\mu^2$.
    For the moment, we consider the $\mu^2=0$ limit of unbroken $Z_2$ symmetry.   The width is
\be
\Gamma(H^0 \ra h^0h^0) = \frac{\lambda_{Hhh}^2}{8\pi M_H} \left( 1 - \frac{4m_h^2}{M_H^2}\right)^{1/2}
\ee
where\cite{Kanemura:2004mg}
\be
\lambda_{Hhh} = -\cos(\beta-\alpha)
\biggl(\frac{\sin 2\alpha}{\sin 2\beta}\biggr) \frac{M_{H^0}^2 + 2 M_{h^0}^2}{2v}\, .
\ee
  Since the decay width of $H^0$ into vector 
bosons also depends on $\cos(\beta-\alpha)$, this factor cancels in the branching ratio.
\begin{figure}[tb]
\subfigure[]{
      \includegraphics[width=0.36\textwidth,angle=0,clip]{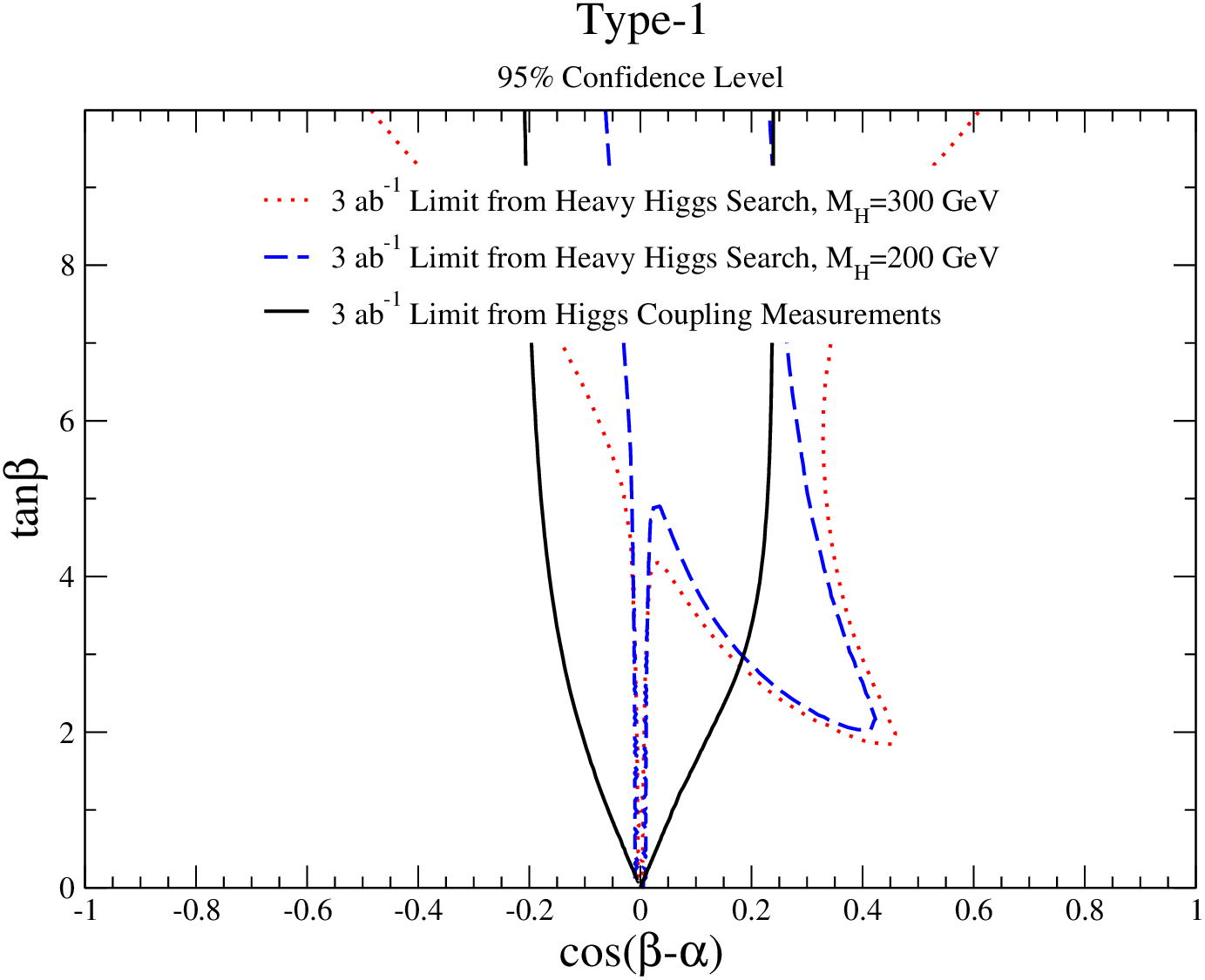}
}
\subfigure[]{
      \includegraphics[width=0.36\textwidth,angle=0,clip]{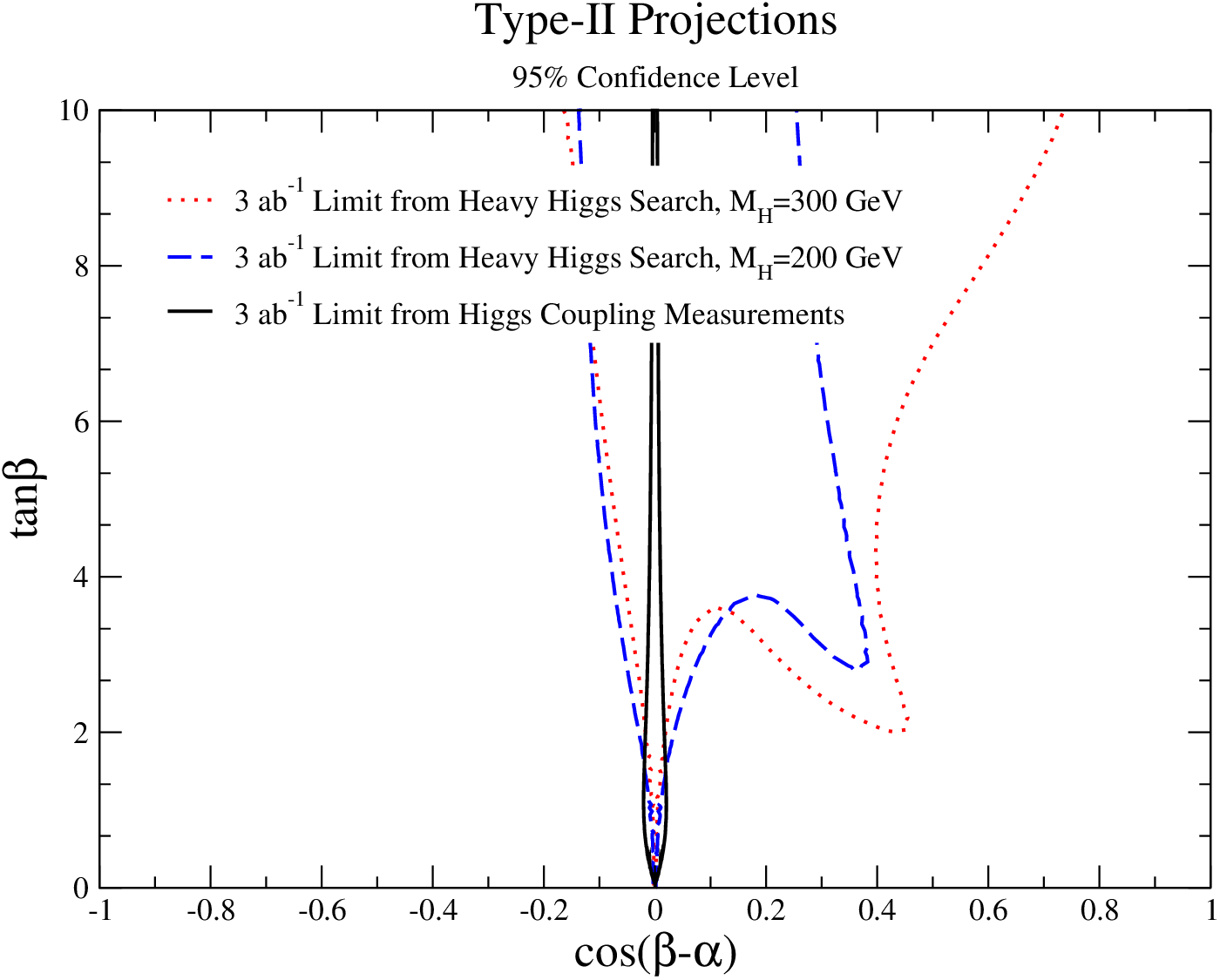}
}
\caption{Allowed regions in the $(\cos(\beta-\alpha),\tan\beta)$ plane in Type I (a), Type II (b) 2HDMs for a potential 
integrated luminosity of $3~ $ab$^{-1}$. The region between the black (solid), blue (dashed), and red (dotted)  lines is allowed at $95\%$ confidence level projected from the Higgs coupling measurements and the heavy Higgs search at $M_{H^0}= 200$ and $300~ GeV$, respectively.
}
\label{fg:MH300}
\end{figure}
The results from the  exclusion of $M_{H^0}=300$ GeV are shown in Figs. \ref{fg:MH23} and \ref{fg:MH300}.   
We see that, as expected due to the opening up of the $H^0 \to h^0h^0$ 
channel,  the exclusion region in the type-I model is smaller than 
from $M_H=200$ GeV, but is still not insubstantial, and becomes quite significant at high integrated luminosity.    In the type-II model, the only additional exclusion regions are at relatively low $\tan\beta$.    Note the dip at $\cos(\beta-\alpha)$ near zero - this occurs because in that limit, 
both $H^0 \to VV$ and $H^0 \to h^0h^0$ vanish,  leaving $H^0 \ra b\bar{b}$ as the dominant decay.   
The results for $M_{H^0}=400$ GeV are not shown.   The additional parameter-space excluded is restricted to a small region for small $\tan\beta$ in the type-I and lepton-specific models.    Clearly, the bounds for higher masses will be weaker.

\begin{figure}[tb]
\subfigure[]{
      \includegraphics[width=0.36\textwidth,angle=0,clip]{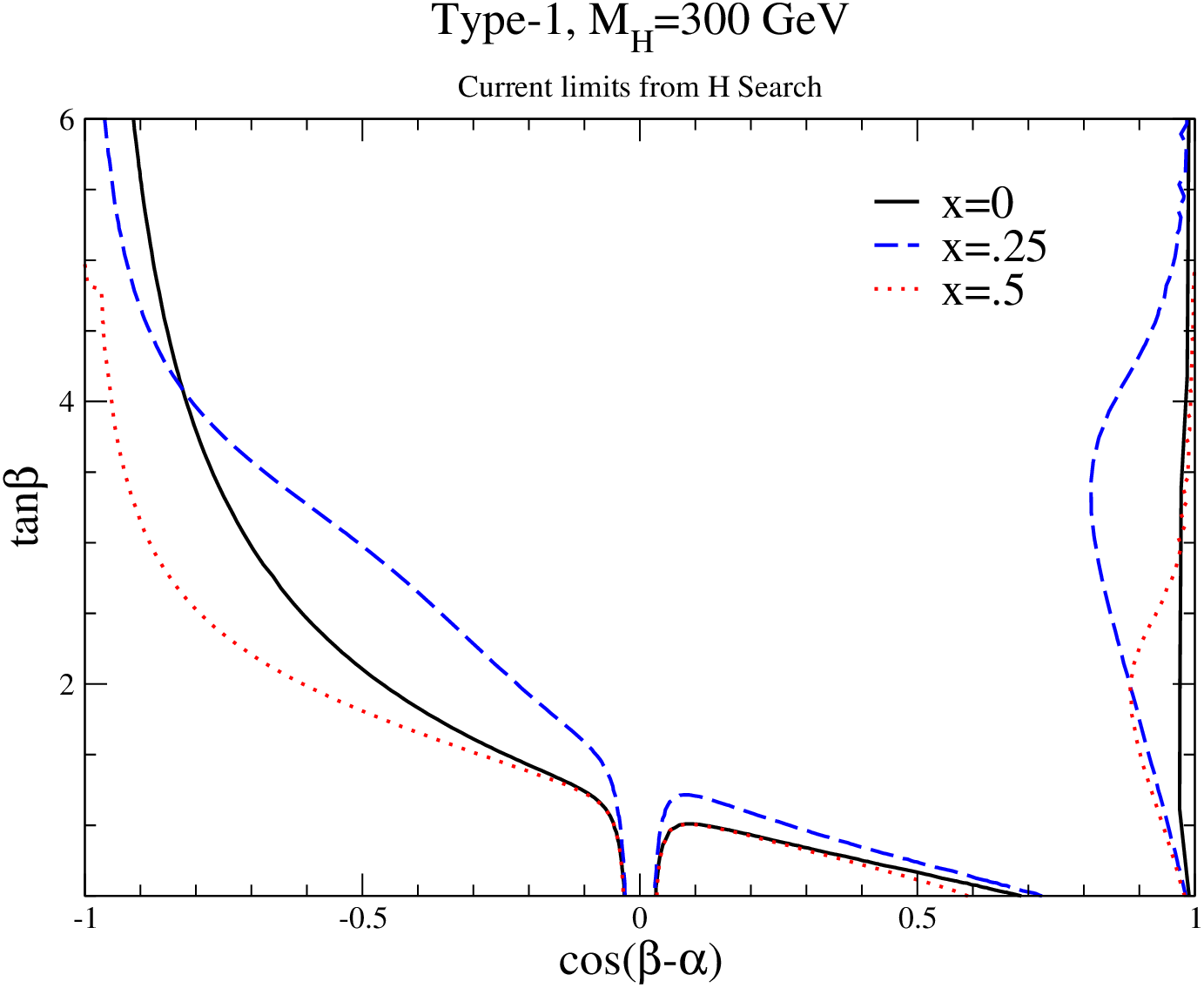}
}
\subfigure[]{
      \includegraphics[width=0.36\textwidth,angle=0,clip]{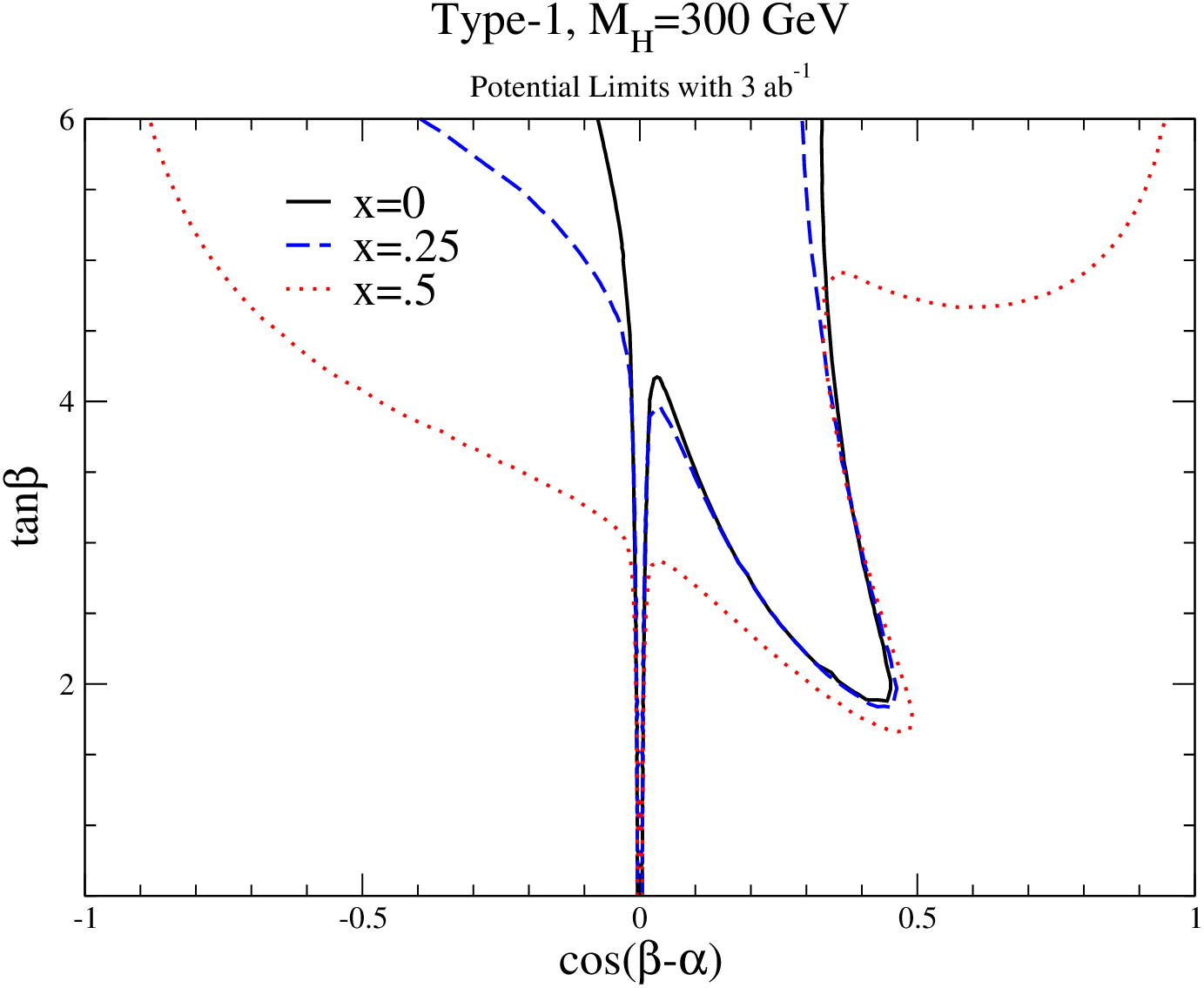}
}
\caption{Allowed regions in the $(\cos(\beta-\alpha),\tan\beta$) plane from the current limits found from the heavy Higgs 
search (a) and the projected limits for an integrated luminosity of $3~ $ab$^{-1}$ (b) in the  Type- I 2HDM with
$ M_{H^0}=300$~ GeV.  
In (a), the regions above the horizontal  black (solid), blue (dashed), and red (dotted)
 lines and to the right of the vertical lines at small $\cos(\alpha-\beta)$  are allowed at $95\%$ confidence level corresponding to $x=0, 0.25$ and $0.5$, respectively.
 In (b), the allowed regions are those above  and enclosed by the curves. }
\label{fg:varyx}
\end{figure}

In the above, we assumed that the $\mu^2$ term, which softly breaks the $Z_2$ symmetry, is absent.   This is technically natural, and in many models the term is naturally small.  If it is not small, however, it will affect our results.  Including the term causes the $Hhh$ coupling to be multiplied\cite{Kanemura:2004mg} by a factor of
\be
\lambda_{Hhh}\rightarrow \lambda_{Hhh}\biggl\{1 - x\left( \frac{3}{\sin 2\beta} - \frac{1}{\sin 2\alpha} \right)\biggr\}\, ,
\ee
where $x \equiv 2\mu^2/(M_{H^0}^2+2M_{h^0}^2)$.   In Fig. \ref{fg:varyx}, we have shown, for the type-I model, how our results are modified as $x$ is varied.

\section{Conclusions}

The discovery of the $125$ GeV Higgs boson and the measurement of its branching ratios has initiated the exploration of the electroweak symmetry breaking sector.   
The implications of the discovery for the simplest extensions of the Standard Model, the two-Higgs-doublet models have
been extensively studied  and the allowed regions of parameter-space determined.
 In this paper,
  we examined the projected sensitivity of these analyses when the LHC has acquired $300~$fb$^{-1}$ and 
  $3000~$fb$^{-1}$ and 
  demonstrated that  LHC bounds on a heavy Standard Model Higgs (between $200$ and $400~ GeV$) can further restrict the parameter-space.   
  In particular, for the type-I 2HDM with a heavy Higgs mass of $200$ GeV, the parameter-space allowed from branching ratios of the $125$ GeV Higgs can
   be shrunk by more than a factor of two by including bounds from the heavy Higgs searches.   It is thus important, in the LHC upgrade, to continue these searches.

{\bf Acknowledgments}

We would like to thank Nathaniel Craig,  Rui Santos, Scott Thomas and Gordon Watts for helpful discussions.   The work of C-Y.C. and S.D. is supported by the United States Department of Energy under Grant DE-AC02-98CH10886 and the work of M.S. is supported by the National Science Foundation under Grant NSF-PHY-1068008.

\vskip -.5in
\begin{table}[tp]
\renewcommand{\arraystretch}{1.4}
\caption{Measured Higgs Signal Strengths}
\centering
\begin{tabular}{|c|c|c|}
\hline
 Decay          & Production & Measured Signal Strength $R^{meas}$ \\ \hline
{$\gamma \gamma$} 
                & ggF        & $1.6^{+ 0.3 +0.3}_{-0.3-0.2}$, [ATLAS] \cite{atlas-13012}\\ 
                & VBF        & $1.7^{+ 0.8 +0.5}_{-0.8-0.4}$  [ATLAS]\cite{atlas-13012}\\ 
                & Vh        & $1.8^{+ 1.5 +0.3}_{-1.3-0.3}$  [ATLAS]\cite{atlas-13012}\\ 
                & inclusive  & $1.65^{+ 0.24 +0.25}_{-0.24-0.18}$   [ATLAS]\cite{atlas-13012}\\ 
                & ggF+tth    & $0.52 \pm 0.5$   [CMS]\cite{cms-13001}\\ 
                & VBF+Vh     & $1.48^{+1.24}_{-1.07}$  [CMS]\cite{cms-13001}\\ 
                & inclusive  & $0.78^{+0.28}_{-0.26}$      [CMS]\cite{cms-13001}\\
                & ggF        & $6.1^{+3.3}_{-3.2}$  [Tevatron]\cite{hcp:Enari}\\ \hline
%
{$W W$}             
                & ggF        & $0.82 \pm 0.36$          [ATLAS] \cite{atlas-13030}\\ 
                & VBF+Vh     & $1.66 \pm 0.79$    [CMS]\cite{atlas-13030}\\ 
                & inclusive  & $1.01 \pm 0.31$   [ATLAS]\cite{atlas-13030}\\ 
                & ggF        & $0.76 \pm 0.21$        [CMS]\cite{cms-13003}\\ 
                & ggF        & $0.8^{+0.9}_{-0.8}$    [Tevatron]\cite{hcp:Enari}\\ \hline
%
{$ZZ$}              
                & ggF        & $1.8^{+0.8}_{-0.5}$   [ATLAS] \cite{atlas-13013}\\ 
                & VBF+Vh     & $1.2^{+3.8}_{-1.4}$   [CMS]\cite{atlas-13013}\\ 
                & inclusive  & $1.5 \pm 0.4$         [ATLAS]\cite{atlas-13013}\\ 
                & ggF        & $0.9^{+0.5}_{-0.4}$   [ATLAS] \cite{cms-13002}\\ 
                & VBF+Vh     & $1.0^{+2.4}_{-2.3}$   [CMS]\cite{cms-13002}\\ 
                & inclusive  & $0.91^{+0.30}_{-0.24}$ [ATLAS]\cite{cms-13002}\\ \hline
\end{tabular}

\vspace{-1ex}
\label{tab:models1}
\end{table}

\begin{table}[tp]
\renewcommand{\arraystretch}{1.4}
\caption{Measured Higgs Signal Strengths}
\centering
\begin{tabular}{|c|c|c|}
\hline
 Decay          & Production & Measured Signal Strength $R^{meas}$ \\ \hline
%
{$b\bar{b}$}      
                & Vh         & $-0.4 \pm 1.0$    [ATLAS] \cite{atlas-170}\\ 
                & Vh         & $1.3^{+0.7}_{-0.6}$      [CMS]\cite{cms-044}\\ 
                & Vh         & $1.56^{+0.72}_{-0.73}$   [Tevatron]\cite{hcp:Enari}\\ \hline
{$\tau^+ \tau^-$}
                & ggF        & $2.4 \pm 1.5$          [ATLAS]\cite{atlas-160}\\ 
                & VBF        & $-0.4 \pm 1.5$         [ATLAS]\cite{atlas-160}\\ 
                & inclusive  & $0.8 \pm 0.7$          [ATLAS]\cite{atlas-170}\\ 
                & ggF        & $0.73 \pm 0.50$        [CMS]\cite{cms-13004}\\ 
                & VBF        & $1.37^{+0.56}_{-0.58}$  [CMS]\cite{cms-13004}\\ 
                & Vh         & $0.75^{+1.44}_{-1.40}$    [CMS]\cite{cms-13004}\\ 
                & inclusive  & $1.1 \pm 0.4$        [CMS]\cite{cms-13004}\\ 
                & ggF        & $2.1^{+2.2}_{-1.9}$    [Tevatron]\cite{hcp:Enari}\\ \hline                 
\end{tabular}

\vspace{-1ex}
\label{tab:models2}
\end{table}

\end{document}